\documentclass[11pt,twoside]{article}

\usepackage{asp2006}
\usepackage{epsf}
\usepackage{lscape}

\begin{document}
\title{The Environmental Impact of Galaxy Evolution}   
\author{Jesper Rasmussen}
\affil{Carnegie Observatories, 813 Santa Barbara Street, Pasadena, CA 91101, 
       USA (Chandra Fellow)}   
\author{Trevor Ponman}  
\affil{School of Physics and Astronomy,
    University of Birmingham, Edgbaston, Birmingham B15 2TT, United Kingdom}

\begin{abstract}
Galaxy evolution reveals itself not only through the evolving
properties of galaxies themselves but also through its impact on the
surrounding environment. The intergalactic medium in particular holds
a fossil record of past galaxy activity, imprinted on its
thermodynamic and chemical properties. This is most easily discerned
in small galaxy groups, where the gravitational heating of this gas
renders it observable by X-ray telescopes while still leaving its
properties highly susceptible to the effects of galactic
feedback. X-ray observations of the hot gas in groups can therefore
provide a view of galactic feedback history that can complement
dedicated studies of AGN and star formation activity at low and high
redshift. Based on high-quality X-ray data of a sample of nearby
groups, we present initial results of such a study and discuss some
implications for the AGN and star formation histories of the group
members.
\end{abstract}

\section{Galaxy Groups as Probes of Cosmic Feedback}

In recent years, deep galaxy surveys such as GOODS \citep{giav04} have
been providing a wealth of high-quality data on the properties of
galaxies across a wide range of redshifts and masses. In combination
with large-area surveys of the low-redshift Universe (e.g.\ the SDSS),
this is enabling detailed studies of the history of stellar mass
assembly, star formation activity, and nuclear activity across
considerable look-back times, providing a much more detailed view of
galaxy evolution than available just a decade ago.

However, a complete understanding of galaxy evolution and the
processes driving it is unlikely to emerge from statistical
considerations applied to large galaxy samples alone. One of several
complimentary approaches is to consider the {\em impact} of galaxy
activity on the surrounding environment. Groups of galaxies provide
particularly useful laboratories for such studies, not only because
they represent a very common galaxy environment in the nearby
Universe, but also because they contain a hot intracluster medium
(ICM) whose properties (thermal pressure, entropy, metallicity) are
highly susceptible to non-gravitational processes such as those
associated with galactic feedback from star formation and AGN
activity. X-ray studies of this hot gas can therefore provide
information on the integrated feedback activity of galaxies. Using a
sample of 15 X-ray bright groups observed with {\em Chandra}, we are
using measurements of the metal distribution in groups to unravel some
of the details of the history of star formation and nuclear activity
in the group members.

\section{Supernova Feedback and Star Formation History}

From the {\em Chandra} data, we have measurements of the radial
distribution of the abundance of iron and silicon in the intragroup
gas for all 15 systems (details of the group sample and data reduction
can be found in \citealt{rasm07}). For an assumed set of supernova
(SN) yields \citep{iwam99,nomo06}, the results for either element can
be uniquely decomposed into contributions from SN~Ia and SN~II and the
results then stacked to provide mean profiles for the entire
sample. In Fig.~\ref{fig,SN} we show the resulting average
contribution of SN~Ia relative to that of SN~II within the sample,
plotted as a function of radius in units of $r_{500}$.
The abundance pattern clearly implies a strong dominance of SN~II
enrichment at large radii, where most of the intragroup gas resides.
\setcounter{figure}{0}
\begin{figure}[htb]
\plotfiddle{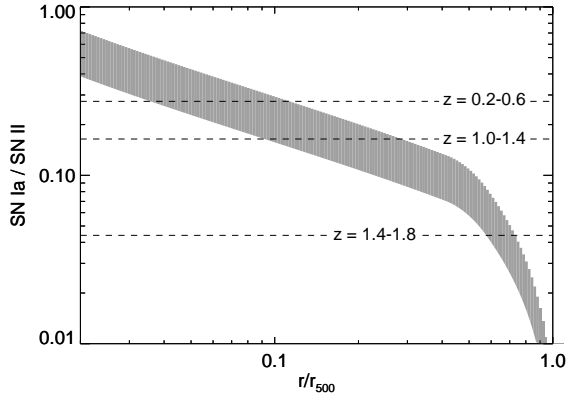}{5.1cm}{0}{42}{42}{-110}{0}
 \caption{The mean number ratio of SN~Ia vs.\ SN~II in our groups as
   inferred from the ICM enrichment pattern, with the shaded region
   representing the typical relative uncertainty of 25\%. For
   comparison, dashed lines show observed and predicted ratios in
   different redshift intervals from deep field data \citep{dahl04}.}
 \label{fig,SN}
 \end{figure}
Comparison to the ratio of SN~Ia vs SN~II measured in the GOODS survey
\citep{dahl04} over a range of redshifts shows that our inferred SN
ratios well outside the group cores, at $r\ga 0.5r_{500}$, are
inconsistent with observed values at low-to-intermediate redshifts ($z
\la 0.6$) but broadly agrees with predictions at $z \ga 1.5$. The
comparison data involve measured SN~Ia rates out to $z\approx 1.8$,
and SN~II rates beyond $z\approx 1$ as predicted from evolutionary
models of the cosmic star formation rate (see Fig.~2 in
\citealt{dahl04}). The immediate implication of Fig.~\ref{fig,SN} is
that most enrichment, and hence SN and star formation activity in the
groups, must have taken place reasonably close to the peak of the
cosmic star formation rate at $z\sim 2-3$.

Further insight may be gained by considering the total metal mass in
the groups generated by each of the two SN types, normalized by the
aggregate $K$-band luminosity $L_K$ of the group members. Using the
adopted SN yields and the SN rate per unit $L_K$ \citep{mann05}
observed in local early-type galaxies (which dominate the optical
output in our groups), these metal mass-to-light ratios can be
translated into enrichment time-scales. The latter will be lower
limits, however, because we do not account for metals locked in stars
or metals ejected beyond $r_{500}$ by galaxy winds, nor for the fact
that $L_K$ must have been smaller in the past due to the continued
growth of stellar mass in group members and the addition of further
members over time.  Fig.~\ref{fig,times} (left) shows the results for
iron from SN~Ia, revealing time-scales in excess of $\ga 10$~Gyr in
many cases and suggesting that SN~Ia at current rates cannot have
produced the required amount of Fe in several of the groups.
\begin{figure}[htb]
\plottwo{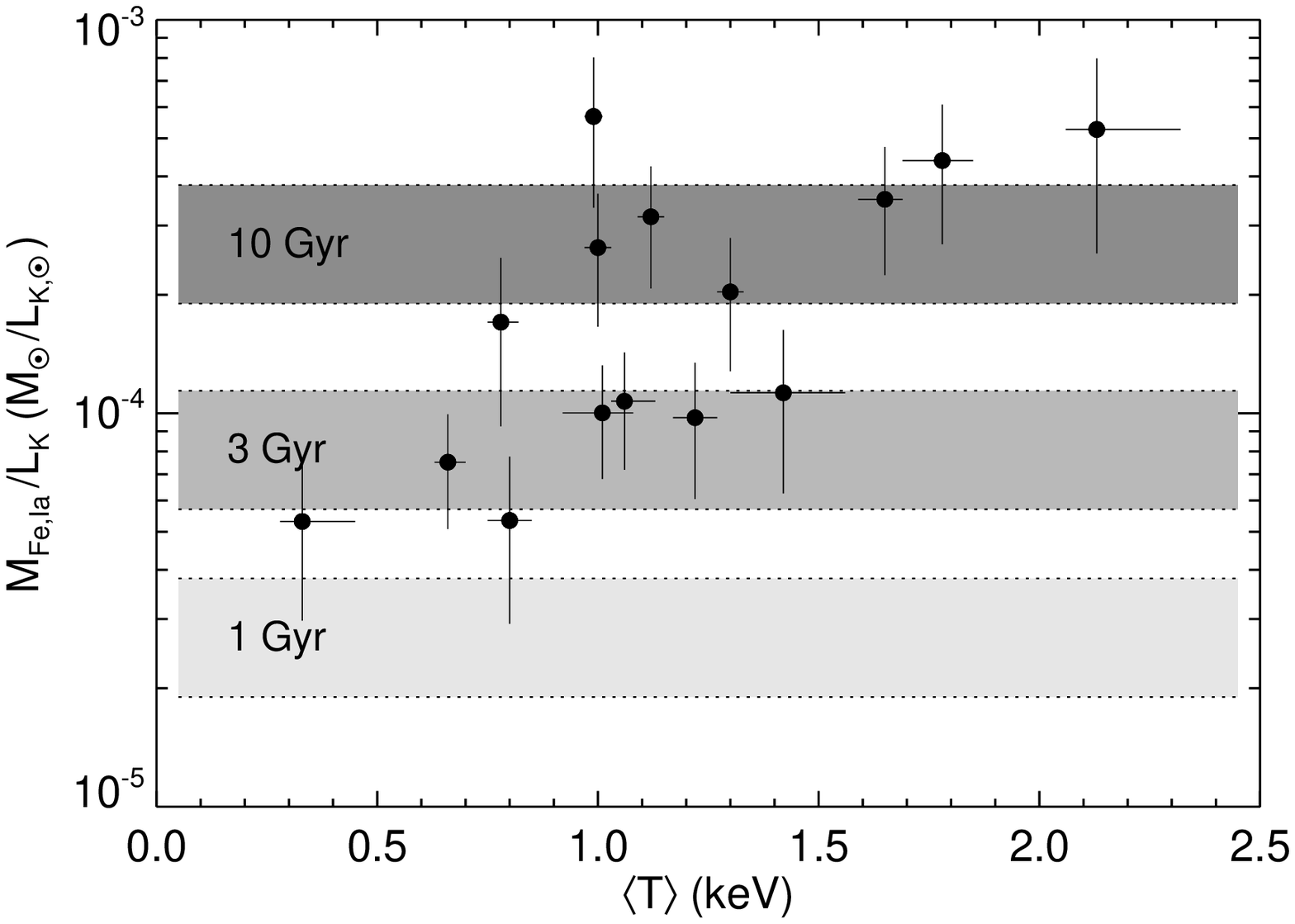}{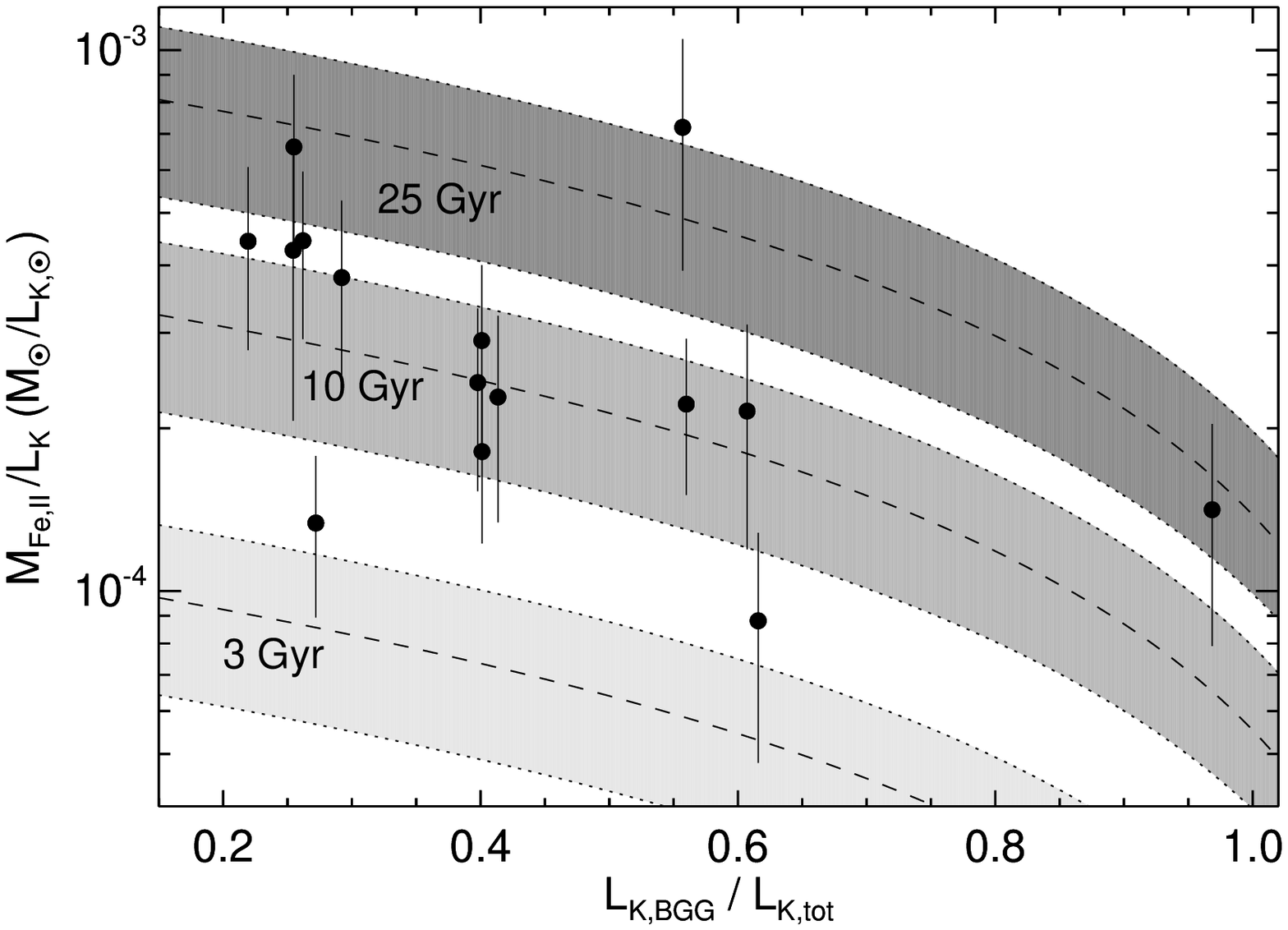}
 \caption{{\itshape Left:\/} $K$-band mass-to-light ratio of iron
   within $r_{500}$ produced by SN~Ia in the groups, as a function of
   mean X-ray temperature. Shaded regions show the corresponding
   time-scales for SN~Ia in the group members to produce the observed
   metal mass, with uncertainties reflecting those of the SN~Ia rates
   in local early-types.  {\itshape Right:\/} The same for SN~II,
   assuming SN rates in local spirals for all satellite galaxies, as a
   function of $K$-band luminosity ratio of the central galaxy to all
   group members.}
\label{fig,times}
 \end{figure}
This issue is even more acute for SN~II, for which only upper limits
to their rate in local early-types are available. Even if, for the
sake of argument, assuming SN~II rates in line with those of nearby
late-type {\em spirals} for all galaxies except the central brightest
group galaxy (BGG), the time-scales are still prohibitively large in
most cases (Fig.~\ref{fig,times} right). Hence, the inferred
enrichment time-scales require much higher specific SN rates in the
past in the group members, independently confirming the
well-established need for a rise in the cosmic star formation rate
density out to at least $z \sim 2-3$, as inferred from galaxy
surveys. While similar results have been reported for massive galaxy
clusters (e.g.\ \citealt{fino00}), this has not previously been tested
at the far more common mass scale of galaxy groups.

\section{Constraining SN and AGN Feedback}

Assuming that energy and metals have been released proportionally from
supernovae to the ICM, the SN ratios and metal masses implied by
Figs.~\ref{fig,SN} and \ref{fig,times} allow us to estimate the total
SN energy imparted to the hot gas for an assumed SN explosion energy
of 10$^{51}$ erg. The resulting values within $r_{500}$, shown in
Fig.~\ref{fig,E} (left), scatter around a mean of $\sim 0.6$~keV per
ICM particle, with no clear trend with group ``mass'' $\langle T
\rangle$. Both the mean and scatter are in broad agreement with
results of hydrodynamical simulations of groups involving
momentum-driven galaxy winds to account for ICM enrichment
\citep{dave08}.

In principle, the inferred SN energies can also help constrain the
impact of AGN feedback in the groups. As a first crude step, one could
argue that the combined energy input from these feedback processes
cannot substantially have exceeded the sum of the ICM thermal energy
and its integrated energy losses without unbinding the hot gas
(according to the virial theorem). Under this assumption, and
evaluating the total radiative energy losses from the ICM on the basis
of its current X-ray luminosity integrated over a 10~Gyr time-scale,
the resulting total allowed AGN heating energy in each group is shown
in Fig.~\ref{fig,E} (right).
\begin{figure}[htb]
\plottwo{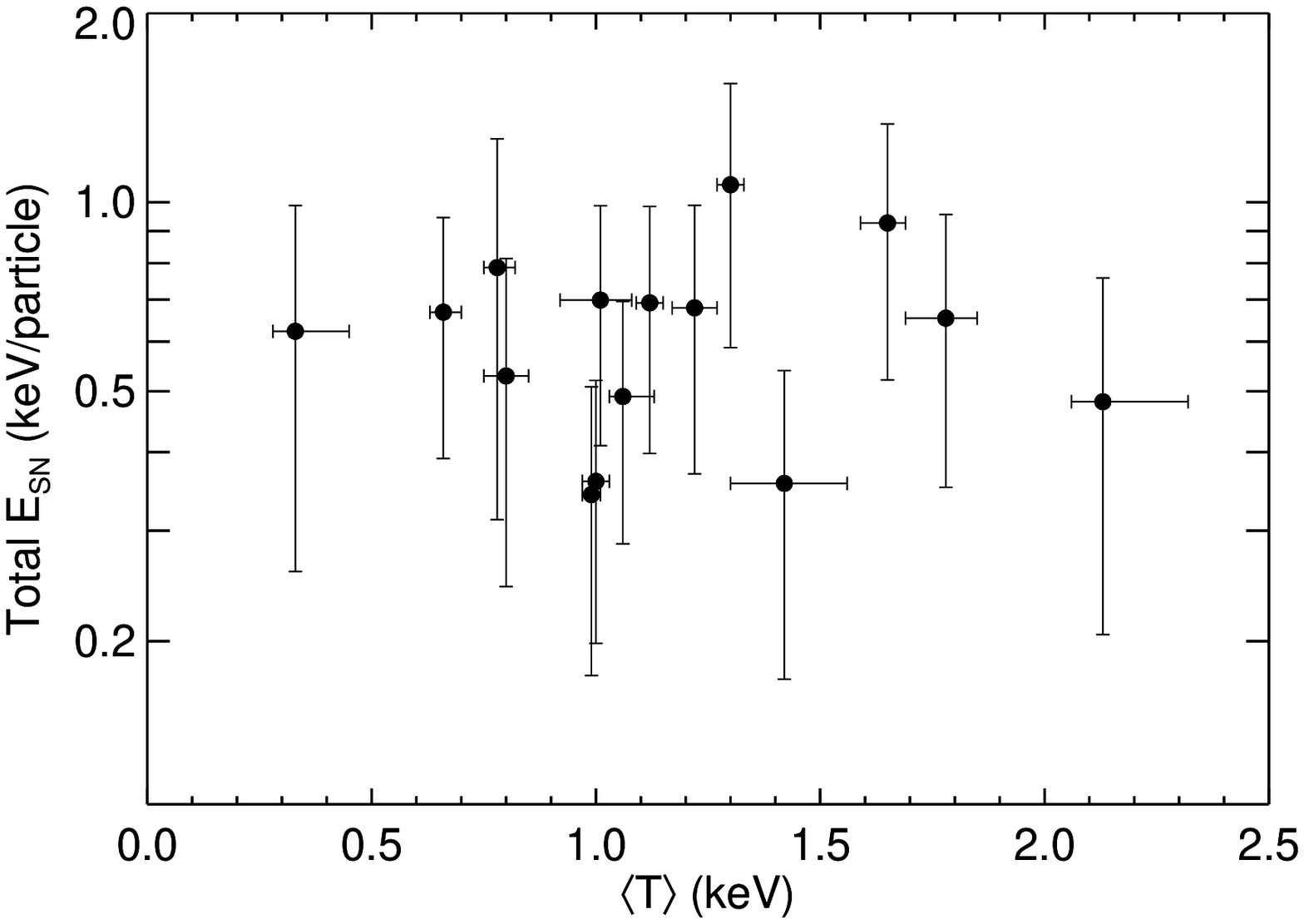}{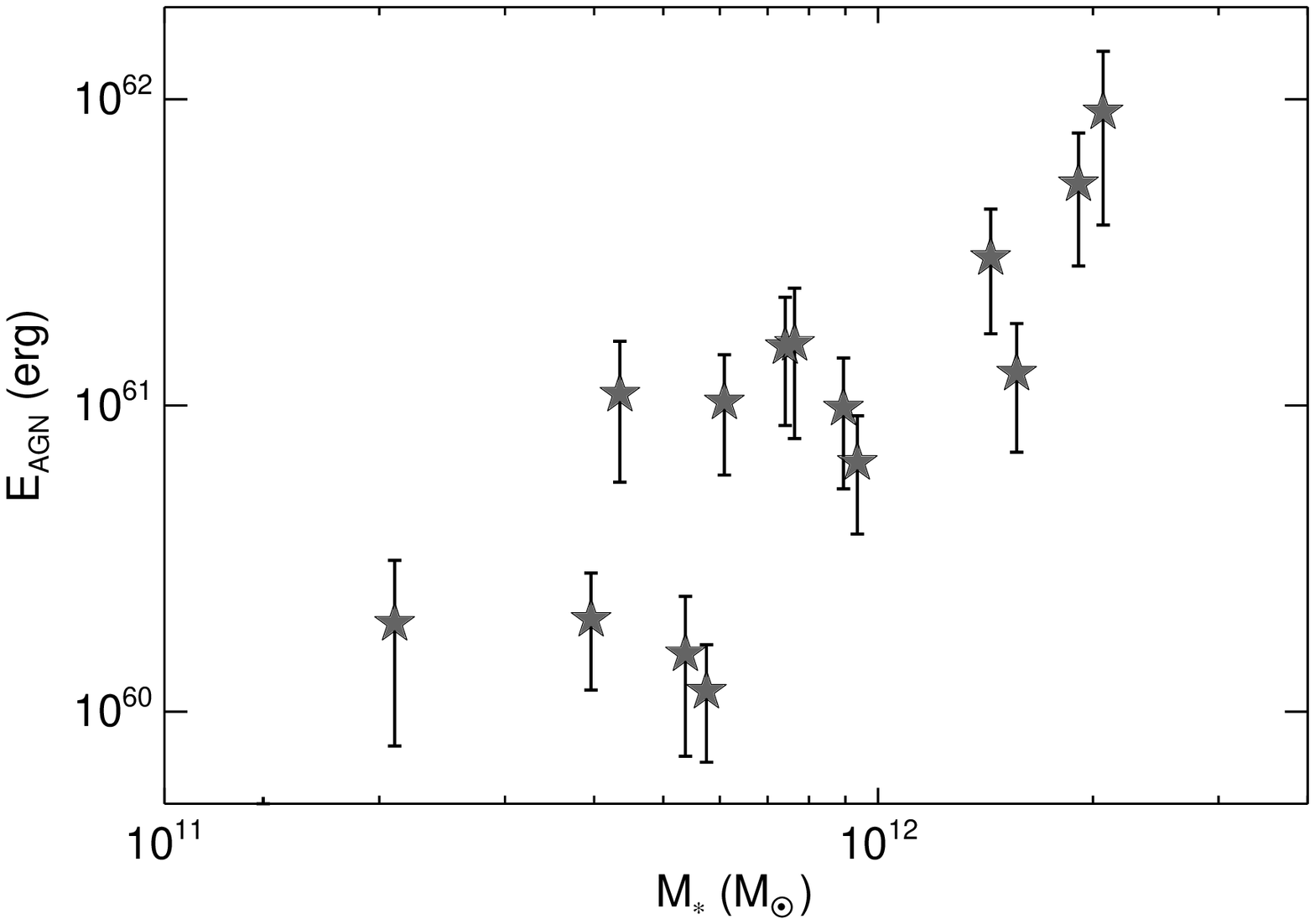}
 \caption{{\itshape Left:\/} SN energy per ICM particle associated
    with the observed ICM metal masses.  {\itshape Right:\/} Total
    allowed AGN heating energy as a function of total stellar mass in
    each group.}
  \label{fig,E}
 \end{figure}
This would suggest that for a typical $T\sim 1$~keV system with a
total stellar mass $M_\ast \sim 1\times 10^{12}$~M$_\odot$, the
integrated AGN heating energy cannot substantially have exceeded $\sim
10^{49}$~erg per M$_\odot$ of stellar mass. To some extent, however,
the above approach mainly probes the resilience of the {\em current}
ICM to additional heating, and it also neglects any bulk kinetic
energy imparted to the gas which may have modified its density
distribution. Such a contribution could be substantial, especially in
the poorest systems, but should more accurately reflect the maximum
energy input that can have occurred in the groups. Hence, more robust
constraints on AGN feedback would arise from assessing the amount of
work done against gravity in establishing current gas mass fractions
and distributions within $r_{500}$ relative to those of massive
clusters.

Assuming that the AGN heating energy in Fig.~\ref{fig,E} has been
released over a 10~Gyr timescale, the resulting time-averaged heating
power in the groups is typically an order of magnitude larger than the
current mechanical AGN luminosity of the BGG, as estimated from its
observed 1.4-GHz radio power and the relation of \citet{birz04}.
Significantly more powerful AGN activity in the past within these
groups is thus allowed, but not necessarily required, by the above
results, in qualitative agreement with the inferred rise in the AGN
luminosity density out to $z \approx 2$ (e.g.\ \citealt{hopk07}). Some
further implications for AGN accretion and supermassive black hole
growth can be obtained by estimating current central black hole masses
$M_{\rm BH}$ from the observed BGG bulge velocity dispersion (e.g.\
\citealt{gebh00}). The energies in Fig.~\ref{fig,E} then provide rough
upper limits to the efficiency $\eta \sim E_{\rm AGN}/(M_{\rm BH}c^2)$
with which mass accreted by a central black hole in the BGG's has been
converted into heating energy in the groups. On average, this number
is $\eta \approx 1-2$\% for our sample, rising to $\eta \approx 5$\%
in the hottest groups.  Although these numbers should clearly be
regarded as tentative at present, it is interesting to note that they
are consistent with estimates of the ratio between black hole
accretion rate and AGN jet power in bright elliptical galaxies
($\approx 2$\%; \citealt{alle06}).

\section{Summary}

Our work demonstrates that the SN and AGN feedback history of galaxies
can be probed by studying how such feedback processes have affected
the hot gas surrounding galaxies in nearby galaxy groups. This
provides a useful complementary approach to those based on large
multi-wavelength galaxy surveys. Specifically, comparison of the
observed amount of metals in the hot intragroup gas to the present-day
optical properties of the group members indicate much higher supernova
and star formation rates per stellar mass in the past. By further
requiring that the observed metal masses be produced within a Hubble
time, these findings could in principle be quantitatively checked
against models predicting the redshift evolution of the specific star
formation rate and stellar mass in galaxies of a given present-day
mass.

Our observations also enable crude constraints on the integrated
impact of AGN feedback from the group members, providing rough upper
limits to the total AGN heating energy released per stellar mass, and
to the efficiency with which central supermassive black holes have
converted accreted mass into heating energy. With some modifications,
our approach could eventually deliver robust constraints on models of
galaxy formation and evolution which include the growth of central
supermassive black holes and the associated AGN feedback.

\acknowledgements JR acknowledges support provided by the National
Aeronautics and Space Administration through Chandra Postdoctoral
Fellowship Award Number PF7-80050.

\end{document}